\documentstyle[12pt]{article}
\newcommand{\sn}{\mbox{sn}}
\newcommand{\cn}{\mbox{cn}}
\newcommand{\dn}{\mbox{dn}}
\makeatletter
\renewcommand\thesection{\@Roman\c@section}
\renewcommand\thesubsection{\thesection.\@arabic\c@subsection}
\makeatother

\begin{document}

\begin{center}
{\Large \bf Elliptic Gaudin models and elliptic KZ Equations}
\vskip.2in  
{\large  Mark D.Gould $^{a}$, Yao-Zhong Zhang $^{a}$ and Shao-You Zhao
$^{a,b}$}
\vskip.2in
{\em $^a$ Center of Mathematical Physics, Department of mathematics, the
university of Queensland,
Brisbane 4072, Australia  
      
 $^b$ P.O. Box 2324, Department of Physics, Nanjing University,
         Nanjing 210093, The People's Republic of China
}
\end{center}
\begin{abstract}

The Gaudin models based on the face-type elliptic quantum groups and 
the $XYZ$ Gaudin models are studied. The Gaudin model Hamiltonians 
are constructed and are diagonalized by using the algebraic Bethe
ansatz method. The corresponding face-type Knizhnik-Zamolodchikov equations 
and their solutions are given.
\end{abstract}

\newcommand{\sect}[1]{\setcounter{equation}{0}\section{#1}}
\renewcommand{\theequation}{\thesection.\arabic{equation}}

\sect{Introduction}
In \cite{Gau76}, Gaudin proposed a quantum integrable model
describing $N$ spin 1/2 particles  with long-range interactions.
The Gaudin type models played an important role in establishing the
integrability of the Seiberg-Witten theory \cite{Sei94,Bra99}, 
and were used as a testing ground
for ideas such as the functional Bethe ansatz and general procedure
of separation of variables \cite{Skl87,Skl96,Skl99}.
They also have direct applications in condensed matter physics.

The algebra associated with the model proposed by Gaudin is the Lie algebra
$\widehat{sl}_2$. Gaudin's work was later generalized by several
authors \cite{Jur89,Hik92,Skl88}
and integrable Gaudin models based on other Lie algebras were
constructed. The $XYZ$ Gaudin model was counstructed and solved in
\cite{Skl96} by means of the algebraic Bethe ansatz.
The boundary Gaudin magnet associated with the spin 1/2
representation of $\widehat{sl}_2$ was investigated by Hikami
\cite{Hik95} using Sklyanin's boundary quantum inverse
scattering method.

The Knizhnik-Zamolodchikov (KZ) equations were first proposed as a set of
differential equations satisfied by correlation functions of
the Wess-Zumino-Witten models \cite{KZ}. The  relation
between the Gaudin magnets and the KZ equations has been studied in many
papers \cite{Babu94,Hik94,Has94,Fei94}. In \cite{Has94}
and \cite{Hik95},  Hikami gave an integral representation for
solutions of the KZ equations by using the results of the periodic and
boundary Gaudin models. 

In the present paper, we  construct elliptic
Gaudin models based on the face-type elliptic quantum group
$E_{\tau,\eta}(\widehat{sl}_2)$ \cite{fe1} and
boundary elliptic quantum group
${\cal B}E_{\tau,\eta}(\widehat{sl}_2)$ \cite{Fan97} and the boundary $XYZ$
Gaudin models. We diagonalize them by means of the algebraic Bethe
ansatz method. Moreover we construct the face-type elliptic KZ equations 
and their solutions using the off-shell Bethe ansatz equations of the
face-type elliptic Gaudin models. 

This paper is organized as follows. In section 2, we review 
the elliptic quantum group $E_{\tau,\eta}(\widehat{sl}_2)$ 
and the boundary elliptic
quantum group ${\cal B}E_{\tau,\eta}(\widehat{sl}_2)$. Then in section 3, we
construct the corresponding elliptic quantum Gaudin models. Solutions
of the KZ equations based on these Gaudin magnets are also given. In section
4, we give the Hamiltonians,  eigenvalues and Bethe ansatz equations
of the boundary $XYZ$ Gaudin model.
In the last section, we present some discussions.

\sect{Elliptic quantum groups and Bethe ansatz}
In this section we review the  elliptic quantum group
$E_{\tau,\eta}(\widehat{sl}_2)$ \cite{fe1,fe2} and the boundary elliptic 
quantum group ${\cal B}E_{\tau,\eta}(\widehat{sl}_2)$ \cite{Fan97}. 
These elliptic quantum
groups are algebraic structures underlying the 
(boundary) dynamical Yang-Baxter equation in statistical mechanics
and the KZ equations on a torus. 

\subsection{$E_{\tau,\eta}(\widehat{sl}_2)$ and its Bethe ansatz}
Let $h$ be the generator of an one-dimensional commutative Lie algebra,  
and $a(\lambda,w)$, $b(\lambda,w)$, $c(\lambda,w)$, $d(\lambda,w)$
be the elements of the matrix $L(\lambda,w)$. Then
$E_{\tau,\eta}(\widehat{sl}_2)$ is generated by meromorphic functions of the
variable $h$ and the elements of  $L(\lambda,w)$ with non-commutative
entries, subject to the dynamical Yang-Baxter equation (YBE) \cite{fe1}
\begin{eqnarray}
&&R^{(12)}(\lambda-2\eta h,w_{12})
      L^{(1)}(\lambda,w_1)L^{(2)}(\lambda-2\eta h^{(1)},w_2)\nonumber\\
&=& L^{(2)}(\lambda,w_2)L^{(1)}(\lambda-2\eta h^{(2)},w_1)
  R^{(12)}(\lambda,w_{12}). \label{DYBR}
\end{eqnarray} 
Here $w_{ij}=w_i-w_j$ with $w_i$ being spectral parameters, 
$\eta$ is the crossing parameter;
$L^{(i)}(\lambda-2\eta h^{(j)},w_i)$ acts on the $i$-th auxiliary
space and $h^{(j)}$ on the $j$ -th auxiliary space $V=C^2$ by a
diagonal matrix $diag(1,-1)$. $R^{(12)}(\lambda-2\eta h,w_{12})$
acts on the space $C^2\otimes C^2$ with $h$ acting on the quantum space.
The $R$-matrix  is given by 
\begin{eqnarray}
R(\lambda,w)&=&E_{0,0}\otimes E_{0,0}+E_{1,1}\otimes E_{1,1}
               +\alpha(\lambda,w)E_{0,0}\otimes E_{1,1}
               +\alpha(-\lambda,w)E_{1,1}\otimes E_{0,0}\nonumber\\
&&\mbox{} +\beta(\lambda,w)E_{0,1}\otimes E_{1,0}
       +\beta(-\lambda,w)E_{1,0}\otimes E_{0,1}
\end{eqnarray}
with 
\begin{eqnarray*}
&& \alpha(\lambda,w)=\frac{\theta(w)\theta(\lambda+2\eta)}
                          {\theta(w-2\eta)\theta(\lambda)},\quad
   \beta(\lambda,w)=\frac{\theta(\lambda+w)\theta(2\eta)}
                          {\theta(w-2\eta)\theta(\lambda)},\\
&&\theta(\lambda)\equiv\theta(\lambda,\tau)
   =-\sum_{j\in Z}e^{\pi i(j+1/2)^2\tau+2\pi i(j+1/2)(\lambda+1/2)}.
\end{eqnarray*}

The commutation relations between the generators of 
$E_{\tau,\eta}(\widehat{sl}_2)$
are defined by eq.(\ref{DYBR}), and the commutation
relations of the generators with some functions $f(\lambda,h)$,
$g(\lambda,h)$ are \cite{fe1}
\begin{eqnarray}
f(\lambda,h)f(\lambda,h)=g(\lambda,h)f(\lambda,h),&&
f(\lambda-2\eta,h) a(\lambda, w)= a(\lambda,w)f(\lambda,h),\nonumber\\
f(\lambda+2\eta,h) d(\lambda, w)= d(\lambda,w)f(\lambda,h),&&
f(\lambda+2\eta,h+2) b(\lambda, w)= b(\lambda, w)f(\lambda,h),\nonumber\\
f(\lambda-2\eta,h-2) c(\lambda, w)
  = c(\lambda, w)f(\lambda,h).&&
\end{eqnarray}

For an even number $\Lambda\ge 0$ and a complex number $z$, we can
define an evaluation
module $V_{\Lambda}(z)$ of $E_{\tau,\eta}(\widehat{sl}_2)$. Let $e_k$, $k\in
Z_{\ge 0}$ be a set of bases of $V_{\Lambda}(z)$. The action is
defined by 
\begin{eqnarray}
&&f(h)e_k=f(\Lambda-2k)e_k,\quad 
   a(\lambda,w)e_k=g(a,\lambda,w,k)e_k, \nonumber\\
&&b(\lambda,w)e_k=g(b,\lambda,w,k)e_{k+1},\quad
   d(\lambda,w)e_k=g(d,\lambda,w,k)e_k, \nonumber\\
&&c(\lambda,w)e_k=g(c,\lambda,w,k)e_{k-1},\label{repres-per}
\end{eqnarray}
with
\begin{eqnarray}
g(a,\lambda,w,k)&=&\frac{\theta(z-w+(\Lambda+1-2k)\eta)
                         \theta(\lambda+2k\eta)}
                        {\theta(z-w+(\Lambda+1)\eta)
                         \theta(\lambda)},\nonumber\\
g(b,\lambda,w,k)&=&-\frac{\theta(-\lambda+z-w+(\Lambda-1-2k)\eta)
                         \theta(2\eta)}
                        {\theta(z-w+(\Lambda+1)\eta)
                         \theta(\lambda)},\nonumber\\
g(c,\lambda,w,k)&=&-\frac{\theta(-\lambda-z+w+(\Lambda+1-2k)\eta)
                         \theta(2(\Lambda+1-k)\eta)\theta(2k\eta)}
                        {\theta(z-w+(\Lambda+1)\eta)
                         \theta(\lambda)\theta(2\eta)},\nonumber\\
g(d,\lambda,w,k)&=&\frac{\theta(z-w+(-\Lambda+1+2k)\eta)
                         \theta(\lambda-2(\Lambda-k)\eta)}
                        {\theta(z-w+(\Lambda+1)\eta)   
                         \theta(\lambda)}.\label{**}
\end{eqnarray}
For any finite-dimensional module 
$W=V_{\Lambda}$ of $E_{\tau,\eta}(\widehat{sl}_2)$, the transfer
matrix can be defined by
\begin{equation}
t(\lambda,w)\equiv tr_0 L(\lambda,w))=a(\lambda,w)+d(\lambda,w).
                      \label{trans-elli}
\end{equation}
Then $t(\lambda,w)$
preserves the space $H\equiv Fun(W)[0]$ of functions with values in the
zero weight space and commutes pairwise on
$H$, i.e. $t(\lambda,w)t(\lambda,u)=t(\lambda,u)t(\lambda,w)$ on $H$.

Let $W=V_{\Lambda_1}(z_1)\otimes\cdots\otimes V_{\Lambda_N}(z_N)$ be a
tensor product of evaluation modules of $E_{\tau,\eta}(\widehat{sl}_2)$ and let
$\Lambda=\Lambda_1+\cdots+\Lambda_N$. 
Then, the highest weight state of the space $H$ with $W[\Lambda]=Cv_0$ obeys
the following highest weight condition, 
\begin{eqnarray}
&& c(\lambda,w)v_0=0,\quad 
   a(\lambda,w)v_0=v_0,   \nonumber\\
&& d(\lambda,w)=\frac{\theta(\lambda-2\eta\Lambda)}
                             {\theta(\lambda)}
                        \prod_{j=1}^N 
                        \frac{\theta(w-z_j-(-\Lambda+1)\eta)}
                             {\theta(w-z_j-(\Lambda+1)\eta)}v_0.
\end{eqnarray}

Under the framework of the algebraic Bethe ansatz, the Bethe state is
defined by
\begin{equation}
\Phi(t_1,\cdots,t_M)=b(t_1)b(t_2)\cdots b(t_M)v
     \equiv b(t_1)b(t_2)\cdots b(t_M)g(\lambda)v_0,
\end{equation}
where $g(\lambda)\ne 0$ is a meromorphic function. 

According to Felder and Varchenko \cite{fe1,fe2}, $W$ is a
highest weight module of $E_{\tau,\eta}(\widehat{sl}_2)$ with 
$\Lambda=2M\in 2Z_{\ge 0}$. Let
$v(\lambda)=\prod_{j=1}^M\theta(\lambda-2\eta j)$.  Then applying the transfer
matrix (\ref{trans-elli}) to the Bethe state gives rise to
\begin{eqnarray}
&&t(\lambda,w)\Phi(t_1,\cdots,t_M)   \nonumber\\
   &=&(a(\lambda,w)+d(\lambda,w))\Phi(t_1,\cdots,t_M)
                                     \nonumber\\   
&=&\epsilon(\lambda, w)\Phi(t_1,\cdots,t_M)
   +\sum_{j=\alpha}^M x F_\alpha
     \Phi_\alpha(t_1,\cdots,t_{\alpha-1},w,t_{\alpha+1},t_M),
                     \label{trans-form}
\end{eqnarray}
where $F_\alpha$ is a function of $t_\alpha$,
$x=\theta(\lambda+t_\alpha-w)\theta(2\eta)/
   (\theta(t_\alpha-w)\theta(\lambda-2\eta))$
and 
 $$\Phi_\alpha(t_1,\cdots,t_{\alpha-1},w,t_{\alpha+1},t_M)= 
   b(\lambda,t_1)\cdots b(\lambda,t_{\alpha-1})
                              b(\lambda,w)
   b(\lambda,t_{\alpha+1})\cdots b(\lambda,t_M)v.$$ 
It follows  that the Bethe state is an eigenstate of the transfer
matrix $t(\lambda,w)$ if $F_\alpha=0$. Therefore the
eigenvalue of the transfer matrix is
\begin{eqnarray}
\epsilon(\lambda,w)=\prod_{\alpha=1}^M
           \frac{\theta(t_\alpha-w-2\eta)}{\theta(t_\alpha-w)}
    +\prod_{\alpha=1}^M\frac{\theta(t_\alpha-w+2\eta)}{\theta(t_\alpha-w)} 
    \prod_{k=1}^N\frac{\theta(w-z_k-(-\Lambda_k+1)\eta)}
                      {\theta(w-z_k-(\Lambda_k+1)\eta)}\nonumber\\
\end{eqnarray}
with $t_\alpha$ determined by the Bethe ansatz equations,
\begin{equation}
F_\alpha=1-\prod_{\beta\ne\alpha}^M
         \frac{\theta(t_\beta-t_\alpha-2\eta)}
              {\theta(t_\beta-t_\alpha+2\eta)}
    \prod_{k=1}^N\frac{\theta(t_\alpha-z_k-(-\Lambda_k+1)\eta)}
                      {\theta(t_\alpha-z_k-(\Lambda_k+1)\eta)}=0.
   \label{elli-BAE}
\end{equation} 

\subsection{${\cal B}E_{\tau,\eta}(\widehat{sl}_2)$ and its Bethe ansatz}
The boundary elliptic quantum group ${\cal B}E_{\tau,\eta}(\widehat{sl}_2)$ is
generated by $h$ and the 
elements of the matrix ${\cal L}(\lambda,w)\in$
End$(C^2)$, ${\cal A}(\lambda,w)$, ${\cal B}(\lambda,w)$, 
${\cal C}(\lambda,w)$ and ${\cal D}(\lambda,w)$ with non-commutative
entries, subject to the relations \cite{Fan97} 
\begin{eqnarray}
&&R_{21}(\lambda,w_1-w_2){\cal L}_1(\lambda-2\eta h^{(2)},w_1)
R_{12}(\lambda,w_1+w_2){\cal L}_2(\lambda-2\eta h^{(1)},w_2) \nonumber\\
&&\quad={\cal L}_2(\lambda-2\eta h^{(1)},w_2)R_{21}(\lambda,w_1+w_2)
{\cal L}_1(\lambda-2\eta h^{(2)},w_1)R_{12}(\lambda,w_1-w_2).
          \label{re-ellip}  \nonumber\\
\end{eqnarray}

Let $V_\Lambda(z)$ be an evaluation module of  ${\cal
B}E_{\tau,\eta}(\widehat{sl}_2)$ with the basis
$e_k,$ $k\in Z_{\ge 0}$ . This module can be identified with 
the original representation space of $E_{\tau,\eta}(\widehat{sl}_2)$. 
On the module $V_\Lambda(z)$, the
representation of ${\cal B}E_{\tau,\eta}(\widehat{sl}_2)$ is given by
\begin{eqnarray}
f(h)e_k&=&f(\Lambda-2k)e_k,\\[3mm]
{\cal A}(\lambda,w)e_k&=&\{Y(\lambda+2\eta,-w)g(a,\lambda+2\eta,w,k)
     g(d,\lambda+2\eta,-w,k-1) \nonumber\\
    &&\mbox{}-K(\lambda-2\eta(\Lambda-2k+2),w)_{22}
              Y(\lambda+2\eta,-w)\nonumber\\
    &&\mbox{}\times g(c,\lambda+2\eta,w,k)
                    g(a,\lambda+2\eta,-w,k-1)\}e_k,\\[3mm]
{\cal B}(\lambda,w)e_k&=&\{Y(\lambda+2\eta,-w)g(b,\lambda+2\eta,w,k)
     g(d,\lambda+2\eta,-w,k) \nonumber\\
    &&\mbox{}-K(\lambda-2\eta(\Lambda-2k),w)_{22}
              Y(\lambda+2\eta,-w)\nonumber\\
    &&\mbox{}\times g(d,\lambda+2\eta,w,k)
                    g(b,\lambda+2\eta,-w,k)\}e_{k+1},\\[3mm]
{\cal C}(\lambda,w)e_k&=&\{K(\lambda-2\eta(\Lambda-2k+2),w)_{22}
      Y(\lambda-2\eta,-w)X(\lambda-2\eta,k-1)\nonumber\\
    &&\mbox{}\times
         g(c,\lambda-2\eta,w,k)g(a,\lambda-2\eta,-w,k)\nonumber\\
    &&\mbox{}-Y(\lambda-2\eta,-w)X(\lambda-2\eta,k-1)\nonumber\\
    &&\mbox{}\times
         g(a,\lambda-2\eta,w,k)g(c,\lambda-2\eta,-w,k)\}e_{k-1},\\[3mm]
{\cal D}(\lambda,w)e_k&=&\{K(\lambda-2\eta(\Lambda-2k),w)_{22}
      Y(\lambda-2\eta,-w)X(\lambda-2\eta,k)\nonumber\\
    &&\mbox{}\times 
         g(d,\lambda-2\eta,w,k)g(a,\lambda-2\eta,-w,k+1)\nonumber\\
    &&\mbox{}-Y(\lambda-2\eta,-w)X(\lambda-2\eta,k)\nonumber\\
    &&\mbox{}\times
         g(b,\lambda-2\eta,w,k)g(c,\lambda-2\eta,w,k+1)\}e_k,
\end{eqnarray}
where $\xi$ is an arbitrary parameter, $g(i,\lambda,w,k)$,~ $i=a,b,c,d$,
are given by (\ref{**}),
\begin{eqnarray*}
X(\lambda,k)&=&\frac{\theta(\lambda-2(\Lambda+1-2k)\eta)}{
                    \theta(\lambda-2(\Lambda-1-2k)\eta)},\\ 
Y(\lambda,w)&=&\frac{\theta(z-w+(\Lambda+1)\eta)\theta(\lambda)}{
                    \theta(z-w-(\Lambda+1)\eta)
                     \theta(\lambda-2(\Lambda+1)\eta)},
\end{eqnarray*}
and $K(\lambda,w)_{22}$ is the (2,2) element of the K-matrix
\begin{equation}
K(\lambda,w)=\mbox{diag}\left(
        1, \frac{\theta(w+\xi)\theta(w+\lambda-\xi)}
                {\theta(w-\xi)\theta(w-\lambda+\xi)}\right)
\end{equation}
which satisfies the boundary dynamical YBE:
\begin{eqnarray}
&&R_{21}(\lambda,w_1-w_2)K_1(\lambda-2\eta h^{(2)},w_1)
R_{12}(\lambda,w_1+w_2)K_2(\lambda-2\eta h^{(1)},w_2) \nonumber\\
&&\quad=K_2(\lambda-2\eta h^{(1)},w_2)R_{21}(\lambda,w_1+w_2) 
K_1(\lambda-2\eta h^{(2)},w_1)R_{12}(\lambda,w_1-w_2).\nonumber\\
\end{eqnarray}

Similar to the $E_{\tau,\eta}(\widehat{sl}_2)$ case, 
for any module $W=V_\Lambda$ of ${\cal B}E_{\tau,\eta}(\widehat{sl}_2)$, 
the boundary transfer matrix is defined by
\begin{equation}
t^b(\lambda,w)=\mbox{tr}_0K^+(\lambda,w){\cal L}(\lambda,w),
                \label{trans-belli}
\end{equation}
where 
\begin{equation}
K^+(\lambda,w)=\mbox{diag}\left(1,
\frac{\theta(\lambda-2\eta)}{\theta(\lambda+2\eta)}
\frac{\theta(w-2\eta-\xi)}{\theta(w-2\eta+\xi)}
\frac{\theta(w-2\eta-\lambda+\xi)}{\theta(w-2\eta+\lambda-\xi)}
                          \right)
\end{equation} 
is a diagonal solution to the dual boundary dynamical YBE,
\begin{eqnarray}
&&R_{21}(\lambda,-w_1+w_2)K^+_1(\lambda-2\eta h^{(2)},w_1)
\tilde R_{12}(\lambda,-w_1-w_2+4\eta)K^+_2(\lambda-2\eta h^{(1)},w_2)
                   \nonumber\\
&&\quad=K^+_2(\lambda-2\eta h^{(1)},w_2)
         \tilde R_{21}(\lambda,-w_1-w_2+4\eta)
            K^+_1(\lambda-2\eta h^{(2)},w_1)\nonumber\\
&&\quad\quad \mbox{}\times R_{12}(\lambda,-w_1+w_2)\label{dual-K}
\end{eqnarray}
with
\begin{eqnarray}
\tilde{R}(\lambda,w)&=&E_{0,0}\otimes E_{0,0}+E_{1,1}\otimes E_{1,1}
       +\alpha(\lambda,w)E_{0,0}\otimes E_{1,1}
       +\alpha(-\lambda,w)E_{1,1}\otimes E_{0,0}\nonumber\\
&&\mbox{}+\frac{\theta(\lambda-2\eta)}{\theta(\lambda+2\eta)}
        \beta(\lambda,w)E_{0,1}\otimes E_{1,0}
       +\frac{\theta(\lambda+2\eta)}{\theta(\lambda-2\eta)}
        \beta(-\lambda,w)E_{1,0}\otimes E_{0,1}.\nonumber\\
\end{eqnarray}
 
The transfer matrix (\ref{trans-belli}) 
preserves the space $H=Fun(W)[0]$ of functions with values in the zero
weight space $W[0]$. Let
$W=V_{\Lambda_1}(z_1)\otimes\cdots V_{\Lambda_N}(z_N)$ be a tensor product
of evaluation modules of ${\cal B}E_{\tau,\eta}(\widehat{sl}_2)$, 
and let $\Lambda=\Lambda_1+\cdots+\Lambda_N$.
The highest weight vector $v_0$ of $H$ with $W[\Lambda]=Cv_0$ satisfies
the highest weight condition:
\begin{eqnarray}
{\cal C}(\lambda,w)v_0&=&0,\quad \quad 
{\cal A}(\lambda,w)v_0=v_0,\nonumber\\
\tilde{\cal D}(\lambda,w)v_0
&=&\frac{\theta(2w)\theta(w-\xi+\lambda-2(\Lambda+1)\eta)
          \theta(w+\xi-2\eta)}
        {\theta(2w-2\eta)\theta(w+\xi-(\lambda-2\Lambda\eta))
          \theta(w-\xi)}\nonumber\\
&&\mbox{}\times\prod_{j=1}^N
    \frac{\theta(z_j+w+(\Lambda_j-1)\eta)\theta(z_j-w-(\Lambda_j-1)\eta)}
         {\theta(z_j+w-(\Lambda_j+1)\eta)\theta(z_j-w+(\Lambda_j+1)\eta)}
                                       v_0,
\end{eqnarray}
where $\tilde{\cal D}(\lambda,w)$ is determined from the relation:
${\cal D}(\lambda,w)=\theta(\lambda)/\theta(\lambda-2\eta)
   \tilde{\cal D}(\lambda,w) +\beta(\lambda-2\eta,2w){\cal A}(\lambda,w)$.
The space $H=Fun(W)$ is spanned by the Bethe state
\begin{equation}
\Phi^b(t_1,\cdots,t_M)={\cal B}(t_1)\cdots {\cal B}(t_M)v_0.
\end{equation}
Applying the boundary transfer matrix to the Bethe state, one obtains
\begin{equation}
t^b(\lambda,w)\Phi^b(t_1,\cdots,t_M)
     =\epsilon(\lambda,w)\Phi^b(t_1,\cdots,t_M)  
     +\sum_{\alpha=1}^MxF_\alpha^b\Phi^b_\alpha
            (t_1,\cdots,t_{\alpha-1},w,t_{\alpha+1},t_M),
\end{equation}
where $$x=\frac{\theta(2\eta)\theta(2t_\alpha)}
               {\theta(\lambda+2\eta)\theta(2t_\alpha-2\eta)}
\left[ \frac{\theta(2w-4\eta)\theta(w+t_\alpha-\lambda)}                 
               {\theta(2w-2\eta)\theta(w+t_\alpha-2\eta)}
      -\frac{\theta(w-t_\alpha-\lambda-2\eta)}
            {\theta(w-t_\alpha)} \right],
$$
\begin{eqnarray}
   \epsilon(\lambda,w)
&=&\prod_{\alpha=1}^M
   \frac{\theta(w+t_\alpha)\theta(w-t_\alpha+2\eta)}
        {\theta(w+t_\alpha-2\eta)\theta(w-t_\alpha)}
  +\frac{\theta(w-2\eta-\xi)\theta(w-\lambda-2\eta+\xi)}
        {\theta(w+\lambda-\xi)\theta(w+\xi)}\nonumber\\ &&\mbox{}\times
   \frac{\theta(2w)\theta(w-2\eta+\xi) 
         \theta(w+\lambda-2(\Lambda+1)\eta-\xi)}
        {\theta(2w-4\eta)\theta(w-\xi) 
         \theta(w-\lambda+2\Lambda\eta+\xi)}\nonumber\\
&&\mbox{}\times\prod_{\alpha=1}^M
   \frac{\theta(w+t_\alpha-4\eta)\theta(w-t_\alpha-2\eta)}
        {\theta(w+t_\alpha-2\eta)\theta(w-t_\alpha)}\nonumber\\
&&\mbox{}\times\prod_{l=1}^N \frac{\theta(z_l+w+(\Lambda_l-1)\eta)
       \theta(z_l-w-(\Lambda_l-1)\eta)} {\theta(z_l+w-(\Lambda_l+1)\eta)
       \theta(z_l-w+(\Lambda_l+1)\eta)},\label{boundary-e}
\end{eqnarray}
and $F^b_\alpha$ obeys the Bethe ansatz equations:
\begin{eqnarray}
    F^b_\alpha
&=&1-\prod_{\beta=1,\ne\alpha}^M
 \frac{\theta(t_\alpha-t_\beta-2\eta)\theta(t_\alpha+t_\beta-4\eta)}
      {\theta(t_\alpha+t_\beta)\theta(t_\alpha-t_\beta+2\eta)} \nonumber\\
&&\mbox{}\times \frac{\theta(t_\alpha-2\eta-\xi)
       \theta(t_\alpha-2\eta-\lambda+\xi)}
      {\theta(t_\alpha+\lambda-\xi)\theta(t_\alpha+\xi)} \nonumber\\
&&\mbox{}\times \frac{\theta(t_\alpha-2\eta+\xi)
       \theta(t_\alpha-2(\Lambda+1)\eta+\lambda-\xi)}
      {\theta(t_\alpha-\lambda+2\Lambda\eta+\xi)\theta(t_\alpha-\xi)}  
                                \nonumber\\
&&\mbox{}\times\prod_{l=1}^N \frac{\theta(z_l+t_\alpha+(\Lambda_l-1)\eta)
       \theta(z_l-t_\alpha-(\Lambda_l-1)\eta)}
      {\theta(z_l+t_\alpha-(\Lambda_l+1)\eta)
       \theta(z_l-t_\alpha+(\Lambda_l+1)\eta)}\nonumber\\
&=&0.
\end{eqnarray}

\sect{Gaudin magnets and KZ equations based on the elliptic quantum groups}
\subsection{Periodic Gaudin magnet}

The Gaudin model was proposed as a quantum integrable system with
long-range interactions \cite{Gau76}. Such a system can be solved by using
the algebraic Bethe ansatz method. By taking the quasi-classical
limit $\eta\rightarrow 0$ of the transfer matrix of the six-vertex model,
Hikami gave the Hamiltonian of the $XXZ$ Gaudin model \cite{Hik94}.

Foe our case, we expand the transfer matrix (\ref{trans-belli})
around the point $\eta=0$ with the parameter $w=z_j$, to obtain
\begin{equation}
t(\lambda,w=z_j)=\frac{2}{\Lambda_j+1}(1+\eta H_j+O(\eta^2)).\label{exp-t}
\end{equation}
One can prove $[H_j,H_k]$=0 since the first
term on the right hand side of (\ref{exp-t})
is a constant. Therefore, $H_j$ gives the
Hamiltonian associated with $E_{\tau,\eta}(\widehat{sl}_2)$.  From 
(\ref{trans-elli}), we find
\begin{eqnarray}
H_j&=& \left.\frac{d t(\lambda,z_j)}{d\eta}\right|_{\eta=0}\nonumber\\
   &=&\sum_{l=1,\ne j}^N\left[
 W_1(z_l,z_j) E_l^-E_j^+
     + W_2(z_l,z_j) E_l^+E_j^- 
      -2h_j\partial_\lambda\right.\nonumber\\
   &&\mbox{}+\left. 
   \frac{1}{2}\varphi(\lambda)(\Lambda_lh_j+h_l)
  +\frac{1}{2}\varphi(z_j-z_l)(h_lh_j-\Lambda_l)\right],\label{ham-elli}
\end{eqnarray}
where $\varphi(x)\equiv \theta'(x)/\theta(x)$, 
\begin{eqnarray}
W_1(x,y)&=&-2\frac{\theta(-\lambda+x-y)\theta'(0)}
           {\theta(x-y)\theta(\lambda)},\nonumber\\
W_2(x,y)&=&-2\frac{\theta(-\lambda-x+y)\theta'(0)}
           {\theta(x-y)\theta(\lambda)},
\end{eqnarray}
and $h$, $E^\pm$ are defined by
\begin{eqnarray}
&& h_l e_k=(\Lambda_l-2k)e_k,\nonumber\\
&&   E_l^-e_k=e_{k+1},\quad 
    E_l^+e_k=k(\Lambda_l+1-k)e_{k-1}.
\end{eqnarray}
We can check that $h$, $E^\pm$ satisfy the commutation relations
\begin{equation}
[h_i,E_j^\pm]=\pm 2\delta_{i,j}E_j^\pm,\quad
[E^+_i,E^-_j]=\delta_{i,j}h_j.
\end{equation}
The eigenvalues $\epsilon_j$ of $H_j$ can be extracted from
\begin{equation}
\epsilon(\lambda,w=z_j)
   =\frac{2}{\Lambda_j+1}(1+\eta\epsilon_j+O(\eta^2)),
\end{equation}
giving 
\begin{eqnarray}
\epsilon_j=-2\Lambda_j\sum_{\alpha=1}^M\varphi(t_\alpha-z_j)
    +(1-\Lambda_j)\sum_{l=1,\ne j}^N
                  \Lambda_l\varphi(z_j-z_l).
\end{eqnarray}
And the constraints for the eigenvalues are
$f_\alpha=0,~\alpha=1,2,\cdots,M$, with
\begin{equation}
f_\alpha=\sum_{\beta\ne\alpha}^M 4\varphi(t_\beta-t_\alpha)
      +2\sum_{l=1}^N\Lambda_l\varphi(t_\alpha-z_l).
\end{equation}

We now define the following equivalent
Hamiltonian ${\cal H}_j$ of the Gaudin model by shifting $H_j$ by some
constants,
\begin{equation}
{\cal H}_j= H_j
           -(1-\Lambda_j)\sum_{l=1,\ne j}^N\Lambda_l\varphi(z_j-z_l).
\end{equation}
The corresponding eigenvalues are
\begin{equation}
{\cal E}_j=-2\Lambda_j\sum_{\alpha=1}^M\varphi(t_\alpha-z_j).
\end{equation}
Then, from (\ref{trans-form}), we obtain the so-called
off-shell Bethe ansatz equations:
\begin{equation}
{\cal H}_j\phi={\cal E}_j\phi+\sum_{\alpha=1}^M
                       f_\alpha W_1(z_j,t_\alpha)E_j^-\phi_\alpha,
\end{equation}
where $\phi, \phi_\alpha$ are Bethe states for the Gaudin model
given by
\begin{eqnarray}
\displaystyle
\phi&\equiv&\left.\prod_{\alpha=1}^M
   \frac{db(\lambda,t_\alpha)}{d\eta}\right|_{\eta=0}|0>,\nonumber\\
\phi_\alpha&\equiv&\left.\prod_{\beta=1,\ne\alpha}^M
    \frac{db(\lambda,t_\beta)}{d\eta}\right|_{\eta=0}|0>.
\end{eqnarray} 
By using the representation (\ref{repres-per}) of 
$E_{\tau,\eta}(\widehat{sl}_2)$,
we derive the precise form of the Bethe states
\begin{eqnarray}
\phi&=&\prod_{\alpha=1}^M\sum_{k=1}^N
W_1(z_k,t_\alpha)E_k^-|0>,\nonumber\\
\phi_\alpha&=&\prod_{\beta\ne\alpha}^M\sum_{k=1}^N 
   W_1(z_k,t_\beta)E_k^-|0>.
\end{eqnarray}

\subsection{Elliptic KZ equation}
As integrable differential equations,  the KZ equations take the form
\begin{equation}
\nabla_j\Psi=0 \quad \quad \mbox{for}\ \  j=1,2,\cdots,N, \label{kz-elli}
\end{equation}
where the differential operators $\nabla_j$ are defined by the Gaudin
model Hamiltonian ${\cal H}_j$
\begin{equation}
\nabla_j=\kappa\frac{\partial}{\partial z_j}-{\cal H}_j \label{elli-nabla}
\end{equation}
with $\kappa$ being an arbitrary parameter.  Substituting (\ref{ham-elli})
into (\ref{elli-nabla}), we can check 
\begin{equation}
[\nabla_j,\nabla_k]=0,
\end{equation}
which ensures the integrability of the KZ equations.

The function $\Psi(z)$ can be constructed by the hypergeometric function
$\chi(z,t)$ which is a solution of the following equations
\begin{eqnarray}
&&\kappa\frac{\partial}{\partial z_j}\chi={\cal E}_j\chi, \nonumber\\ 
&&\kappa\frac{\partial}{\partial t_\alpha}\chi=f_\alpha\chi.
\end{eqnarray}
The hypergeometric function $\chi(z,t)$ can be written as
\begin{equation}
\chi(z,t)=\prod_{\beta<\alpha}[\theta(t_\alpha-t_\beta)]^{-4/\kappa}
          \prod_{\alpha=1}^M\prod_{j=1}^N
             [\theta(z_j-t_\alpha)]^{2\Lambda_j/\kappa}.
\end{equation}
With the help of $\chi(z,t)$, we obtain
\begin{equation}
\Psi(z)=\oint_C\cdots\oint_C dt_1\cdots dt_M \chi(t,z)\phi(t,z),
\end{equation}
where the integration path $C$ is a closed contour in the
Riemann surface such that the integrand resumes its initial value.

Substituting the expressions of $\nabla_j$ and $\Psi(z)$ into 
({\ref{kz-elli}), we can show that the KZ equations are satisfied.
The proof is as follows
\begin{eqnarray}
\kappa\frac{\partial}{\partial z_j}\Psi(z)&=&
   \oint_C\cdots\oint_C dt_1\cdots dt_M \left(
       \kappa\frac{\partial \chi}{\partial z_j}\phi
       +\kappa\chi\frac{\partial \phi}{\partial z_j}\right)\nonumber\\
&=& \oint_C\cdots\oint_C dt_1\cdots dt_M \left(
       \chi {\cal E}_j\phi
       +\kappa\chi\frac{\partial \phi}{\partial z_j}\right)\nonumber\\
&=&\oint_C\cdots\oint_C dt_1\cdots dt_M \left[
        \chi{\cal H}_j\phi -\kappa\sum_{\alpha=1}^M
\frac{\partial \chi}{\partial t_\alpha}W_1(z_j,t_\alpha)E^-_j\phi_\alpha
      \right.\nonumber\\ &&\mbox{}\left.
-\kappa\chi\sum_{\alpha=1}^M
\frac{\partial }{\partial t_\alpha}\left(
     W_1(z_j,t_\alpha)E_j^-\phi_\alpha\right)\right] \nonumber\\
&=&\oint_C\cdots\oint_C dt_1\cdots dt_M \left[
        {\cal H}_j\chi\phi-\kappa\sum_{\alpha=1}^M
         \frac{\partial }{\partial t_\alpha}\left(\chi
       W_1(z_j,t_\alpha)E_j^- \phi_\alpha\right)\right] \nonumber\\
&=&{\cal H}_j\Psi(z),
\end{eqnarray}

\subsection{Boundary Gaudin magnet and KZ equation}

The Gaudin model Hamiltonian for ${\cal B}E_{\tau,\eta}(\widehat{sl}_2)$ is
found to be the form \cite{Hou00}
\begin{eqnarray}
H_{j}^b(\lambda, z_j)&=&\delta(\lambda,z_j,\xi)\nonumber\\
& &+\;\sum_{k=1}^N \left[K_{22}W_1^+(z_k,z_j)E_k^-E_j^+ \right.
+K^+_{22}W_2^+(z_k,z_j)E_k^+E_j^- \nonumber\\
& &\quad-\;\left.\varphi(z_k+z_j)( h_k h_j-\Lambda_k)
+\varphi(\lambda)(2\Lambda_k- h_k -\Lambda_k h_j)\right]\nonumber\\
& &+\;\sum_{k=1,\ne j}^n\left[W_1^-(z_k,z_j)E_k^-E_j^+
+W_2^-(z_k,z_j)E_k^+E_j^- \right.\nonumber\\
& &\quad +\;\left.\varphi(z_k-z_j)(h_k h_j-\Lambda_k)
-\varphi(\lambda)( h_k -\Lambda_k h_j)\right],
\end{eqnarray}
where $E^\pm, h$ and $\varphi$ are the same as in the periodic case,
$K_{22}$ and  $K_{22}^+$ are $(2,2)$-elements of $K$ and $K^+$
respectively, and
\begin{eqnarray}
W_1^\pm(x,y)&=&-2\frac{\theta(-\lambda+x\pm y)\theta'(0)}
               {\theta(x\pm y)\theta(\lambda)},\nonumber\\
W_2^\pm(x,y)&=&2\frac{\theta(\lambda+x\pm y)\theta'(0)}
               {\theta(x\pm y)\theta(\lambda)},\nonumber\\
\delta(\lambda,z_j,\xi)
&=&(1+2h_j)(h_j-1)
[\varphi(z_j-\lambda+\xi)-\varphi(z_j+\lambda-\xi)]\nonumber\\
& &\mbox{} +(h_j-1)
   [2\varphi(\lambda)+\varphi(z_j-\xi)-\varphi(z_j+\xi)]\nonumber\\
& &\mbox{}+h_j(\Lambda_j-1)\left[\varphi(\lambda)
      +\frac{\varphi''(0)}{\varphi'(0)}\right],
\end{eqnarray}

By (\ref{boundary-e}) we find the eigenvalues of the boundary Gaudin
model Hamiltonian $H_j^b$, 
\begin{eqnarray}
\epsilon_j^b&=&\left.\frac{d\epsilon(\lambda,w=z_j)}{d\eta}\right|_{\eta=0}
                       \nonumber\\
&=&\sum_{\alpha=1}^M
   2\Lambda_j[\varphi(z_j+t_\alpha)+\varphi(z_j-t_\alpha)]
  -(1-\Lambda_j^2)\varphi(2z_j) \nonumber\\
&&\mbox{} +\sum_{\l=1,\ne j}^N 
   \Lambda_l(1-\Lambda_j)[\varphi(z_l+z_j)-\varphi(z_l-z_j)]
               \nonumber\\
&&\mbox{} 
  -(1-\Lambda_j)[
\varphi(z_j+\xi)+\varphi(z_j-\xi)
+\varphi(z_j-\lambda+\xi)+\varphi(z_j+\lambda-\xi)]\nonumber\\
\end{eqnarray}
with the constraints $f_\alpha^b=0,~\alpha=1,2,\cdots,M$, where
\begin{eqnarray}
f^b_\alpha&=& 2 \varphi(t_\alpha-\xi)+2\varphi(t_\alpha+\xi)
  +2(1+\Lambda)[\varphi(t_\alpha-\lambda+\xi)
              +\varphi(t_\alpha+\lambda-\xi)]\nonumber\\  
&&\mbox{}+4\sum_{\beta=1,\ne\alpha}^M
   [\varphi(t_\alpha-t_\beta)+\varphi(t_\alpha+t_\beta)] -\sum_{l=1}^N
   2\Lambda_l[\varphi(z_l+t_\alpha)-\varphi(z_l-t_\alpha)].\nonumber\\
\end{eqnarray}
The Bethe states for the boundary Gaudin model are then given by
\begin{eqnarray}
\phi^b&=&\prod_{\alpha=1}^M\left[\sum_{l=1}^N 
       (W_1^-(z_l,t_\alpha) 
        -K_{22}(t_\alpha)W_1^+(z_l,t_\alpha))E^-_l\right]v_0,\nonumber\\
\phi^b_\alpha&=&\prod_{\beta=1,\ne\alpha}^M\left[\sum_{l=1}^N
       (W_1^-(z_l,t_\beta) 
        -K_{22}(t_\beta)W_1^+(z_l,t_\beta))E^-_l\right]v_0. 
\end{eqnarray}

As in the previous subsection, we define the following equivalent
Hamiltonian ${\cal H}^b_j$ of the boundary Gaudin model: 
\begin{eqnarray}
{\cal H}^b_j&=&-H_j^b +\sum_{\l=1,\ne j}^N 2(\Lambda_l)(1-\Lambda_j)
   [\varphi(z_l+z_j)-\varphi(z_l-z_j)]\nonumber\\
\nonumber\\
&&\mbox{} -(1-\Lambda_j)[z_j+\xi)+\varphi(z_j-\xi)
+\varphi(z_j-\lambda+\xi)+\varphi(z_j+\lambda-\xi)].\nonumber\\
\end{eqnarray}
The corresponding eigenvalues are given by
\begin{eqnarray}
{\cal E}^b_j=(1-\Lambda_j^2)\varphi(2z_j)
     -\sum_{\alpha=1}^M
      2\Lambda_j[\varphi(z_j+t_\alpha)+\varphi(z_j-t_\alpha)],
\end{eqnarray}
and  the off-shell Bethe ansatz equations read
\begin{equation}
{\cal H}^b_j\phi^b={\cal E}^b_j\phi^b+\sum_{\alpha=1}^Mf_\alpha^b 
   \left(W_1^-(z_l,t_\alpha)-W_1^+(z_l,t_\alpha)\right)E_j^-\phi^b_\alpha.
\end{equation} 

Thus the KZ equations based on the boundary Gaudin model
are given by
\begin{equation}
\nabla_j\Psi^b=0 \quad \mbox{for } j=1,2,\cdots,N,
\end{equation}
where the differential operators $\nabla_j$ are defined by
\begin{equation}
\nabla_j=\kappa\frac{\partial}{\partial z_j}-{\cal H}^b_j
\end{equation}
with $\kappa$ being an arbitrary parameter. To ensure the integrability of
the KZ equation, we impose
\begin{equation}
[\nabla_j,\nabla_k]=0,
\end{equation}
This requires $\xi\rightarrow \infty$. With this condition, the
Hamiltonian and the Bethe states of the boundary Gaudin model become
\begin{eqnarray}
{\cal H}_{j}^b(\lambda)&=&-\Delta(\lambda,z_j,\xi)-\sum_{k=1}^N\left[
W_1^+E_k^-E_j^+ +W_2^+E_k^+E_j^-\right.  \nonumber\\
& &\quad-\;\left.\varphi(z_k+z_j)( h_k h_j-\Lambda_k)
+\varphi(\lambda)(2\Lambda_k- h_k -\Lambda_k h_j)\right]\nonumber\\
& &-\;\sum_{k=1,\ne j}^n \left[W_1^-E_k^-E_j^+
+W_2^-E_k^+E_j^- \right.\nonumber\\
& &\quad +\;\left.\varphi(z_k-z_j)(h_k h_j-\Lambda_k)
-\varphi(\lambda)( h_k
  -\Lambda_k h_j)\right],\nonumber\\[3mm]
\phi^b&=&\prod_{\alpha=1}^M\left[\sum_{l=1}^N
       (W_1^-(z_l,t_\alpha)  
        -W_1^+(z_l,t_\alpha))E^-_l\right]v_0,\nonumber\\
\phi^b_\alpha&=&\prod_{\beta=1,\ne\alpha}^M\left[\sum_{l=1}^N
       (W_1^-(z_l,t_\beta)
        -W_1^+(z_l,t_\beta))E^-_l\right]v_0,
\end{eqnarray}
where 
\begin{eqnarray*}
\Delta(\lambda)&=&
 \sum_{\l=1,\ne j}^N
   2(\Lambda_l)(1-\Lambda_j) 
   [\varphi(z_l+z_j)-\varphi(z_l-z_j)]\nonumber\\
&&\mbox{}+ [h_j(\Lambda_j-1)]\varphi(\lambda) +
(\Lambda_j+h_j)(\Lambda_j-1)\frac{\varphi''(0)}{\varphi'(0)}.
\end{eqnarray*}  

The function $\Psi^b$ can be computed by using the
following hypergeometric type integral
\begin{equation}
\Psi^b(z)=\oint_C\cdots\oint_C dt_1\cdots dt_M
 \chi^b(t,z)\phi^b(t,z),
\end{equation}
where the integration path $C$ is a closed contour in the
Riemann surface, similar to that described in the previous subsection. The
hypergeometric kernel $\chi^b(t,z)$ is given by
\begin{eqnarray}
\chi^b(t,z)&=&\prod_{j=1}^N
(\theta(2z_j))^{(1-\Lambda_j^2)/2\kappa}
   \prod_{j=1}^N\prod_{\alpha=1}^M
      [\theta(z_j-t_\alpha)\theta(z_j+t_\alpha)]^{-2\Lambda_j/\kappa}
             \nonumber\\ &&\mbox{}\times
   \prod_{\alpha<\beta}^M 
      [\theta(t_\alpha-t_\beta)\theta(t_\alpha+t_\beta)]^{4/\kappa}.
\end{eqnarray}
One can check that the kernel $\chi^b(t,z)$ satisfies the following
equations
\begin{eqnarray}
&&\kappa\frac{\partial}{\partial z_j}\chi^b={\cal E}^b_j\chi, \nonumber\\
&&\kappa\frac{\partial}{\partial t_\alpha}\chi^b=f^b_\alpha\chi.
\end{eqnarray}
Moreover we can show that $\Psi^b$ satisfies the KZ equations, 
\begin{equation}
\nabla_j\Psi^b(z)=0.
\end{equation} 
The proof is as follows.
\begin{eqnarray}
\kappa\frac{\partial}{\partial z_j}\Psi(z)&=&
   \oint_C\cdots\oint_C dt_1\cdots dt_M \left(
       \kappa\frac{\partial \chi^b}{\partial z_j}\phi^b
       +\kappa\chi^b\frac{\partial \phi^b}{\partial z_j}\right)\nonumber\\
&=& \oint_C\cdots\oint_C dt_1\cdots dt_M \left(
       \chi {\cal E}^b_j\phi^b
       +\kappa\chi^b\frac{\partial \phi^b}{\partial z_j}\right)\nonumber\\
&=&\oint_C\cdots\oint_C dt_1\cdots dt_M \left[
        \chi^b{\cal H}^b_j\phi^b \right.\nonumber\\
&&\mbox{}    -\kappa\sum_{\alpha=1}^M
\frac{\partial \chi^b}{\partial t_\alpha}
\left(W_1^-(z_j,t_\alpha)-W_1^+(z_j,t_\alpha)\right)E^-_j\phi^b_\alpha
      \nonumber\\ &&\mbox{}\left.
-\kappa\chi^b\sum_{\alpha=1}^M
\frac{\partial }{\partial t_\alpha}\left(\left(
W_1^-(z_j,t_\alpha)-W^+_1(z_j,t_\alpha)\right)
    E_j^-\phi^b_\alpha\right)\right]
\nonumber\\
&=&\oint_C\cdots\oint_C dt_1\cdots dt_M \left[
        {\cal H}^b_j\chi\phi \right.  \nonumber\\ &&\mbox{}
-\kappa\sum_{\alpha=1}^M\left.
         \frac{\partial }{\partial t_\alpha}\left(\chi^b
      \left( W_1^-(z_j,t_\alpha)- W_1^+(z_j,t_\alpha)\right)
       E_j^- \phi^b_\alpha\right)\right]
    \nonumber\\
&=&{\cal H}_j^b\Psi(z),
\end{eqnarray}

\sect{$XYZ$ Gaudin magnets}
The XYZ chain or eight-vertex model occupies an important
place in the study of integrable systems \cite{Bax}. The periodic $XYZ$
Gaudin model has been constructed and diagonalized in
\cite{Skl96} by means of the algebraic Bethe ansatz method. The 
corresponding KZ equations and their solutions were given in
\cite{Bab99}. Here we write down a more explicit formula for the Bethe states
for the periodic case. Moreover, we present the Hamiltonian,
eigenvalues and (off-shell) Bethe ansatz equations for the
$XYZ$ Gaudin magnet with an integrable boundary.

The Boltzmann weights of the eight-vertex model are given by
\begin{eqnarray}
R(w)&=&E_{00}\otimes E_{00}+E_{11}\otimes E_{11}
  +\kappa\sn(\eta)\sn(w) (E_{01}\otimes E_{01}+E_{10}\otimes E_{10})   
        \nonumber\\
& &\mbox{} +\frac{\sn(w)}{\sn(w+\eta)}
    (E_{00}\otimes E_{11}+E_{11}\otimes E_{00})
  +\frac{\sn(\eta)}{\sn(w+\eta)}
    (E_{01}\otimes E_{10}+E_{10}\otimes E_{01}), \nonumber\\
\end{eqnarray}
where $\sn(w)\equiv \sn(w,\kappa)$ is the Jacobi elliptic function of modulus
$\kappa$, $0\le \kappa\le 1$. Define the corresponding transfer matrix as
\begin{equation}
t(w)=tr_0R_{01}(w-z_1)\cdots R_{0N}(w-z_N)
\end{equation}
for the periodic case, and as
\begin{eqnarray}
t^b(w)&=&tr_0K^+(w,\xi)R_{01}(w-z_1)\cdots R_{0N}(w-z_N)
             K(w,\xi)          \nonumber\\ &&\times
         R_{0N}(w+z_N)\cdots R_{01}(w+z_1)
\end{eqnarray}
for the boundary case,
where $K(u,\xi)=\mbox{diag}(\sn(\xi+u), \sn(\xi-u))/\sn(\xi)$ \cite{inami}
and $K^+(u,\xi)=K(-u-\eta,\xi)$ with $\xi$ being a parameter.

Then, by using the same method as in the previous sections, one can obtain the
Hamiltonians of the $XYZ$ Gaudin models. 
\vskip.1in	
\noindent {\bf (i) The periodic case:}
\begin{eqnarray}
H_j&=&\sum_{k=1,\ne j}^N\frac{1}{\sn(z_j-z_k)}\left[
    (1+\kappa\sn^2(z_j-z_k)\sigma_j^x\sigma_k^x
   +(1-\kappa\sn^2(z_j-z_k)\sigma_j^y\sigma_k^y\right.\nonumber\\
&&\mbox{}\quad\quad\left.
   +\cn(z_j-z_k)\dn(z_j-z_k)(\sigma_j^z\sigma_k^z-1)\right],\label{8v-G}
\end{eqnarray}
where and throughout this paper, ${\rm cn}(u){\rm dn}(u)=d{\rm sn}(u)/du$.

\vskip.1in
\noindent {\bf (ii)The open boundary case:}
\begin{eqnarray}
H^b_j&=&-\frac{1}{2\sn(\xi-z_j)}\left[
        \cn(\xi-z_j)\dn(\xi-z_j)-\frac{\sn(\xi+z_j)}{\sn(2z_j)}\right]
        (\sigma_j^z+1)\nonumber\\
&&\mbox{}
-\frac{1}{2\sn(\xi+z_j)}\left[
         \cn(\xi+z_j)\dn(\xi+z_j)+\frac{\sn(\xi-z_j)}{\sn(2z_j)}\right]
        (\sigma_j^z-1)
\nonumber\\
&&\mbox{}+ \sum_{k=1\ne
j}^N\frac{1}{\sn(z_j-z_k)}\left[\sigma^+_j\sigma^-_k
  +\sigma^-_j\sigma^+_k+\kappa\sn^2(z_j-z_k)(\sigma^+_j\sigma^+_k
  +\sigma^-_j\sigma^-_k)\right.\nonumber\\
&&\ \ \ \ \ \left. +\cn(z_j-z_k)\dn(z_j-z_k)
  \frac{\sigma^z_j\sigma^z_k-1}{2}\right]\nonumber\\
&&+\ \sum_{k=1\ne
j}^N\frac{1}{\sn(z_j+z_k)}\left[\kappa\sn^2(z_j+z_k)\left(
 \frac{\sn(\xi+z_j)}{\sn(\xi-z_j)}\sigma^+_j\sigma^+_k
+\frac{\sn(\xi-z_j)}{\sn(\xi+z_j)}\sigma^-_j\sigma^-_k\right)\right.\nonumber\\
&&\ \ \ \ \ \left. +\frac{\sn(\xi+z_j)}{\sn(\xi-z_j)}\sigma^+_j\sigma^-_k
 +\frac{\sn(\xi-z_j)}{\sn(\xi+z_j)}\sigma^-_j\sigma^+_k
 +\cn(z_j+z_k)\dn(z_j+z_k)
  \frac{\sigma^z_j\sigma^z_k-1}{2}\right].\nonumber\\
  \label{8v-Gb}
\end{eqnarray}

The Hamiltonians (\ref{8v-G}) and (\ref{8v-Gb}) can be diagonalized by 
using the method similar to the previous section. The results are as follows.
\vskip.1in  
\noindent {\bf (i) The periodic case.}

The eigenvalues of (\ref{8v-G}) are found to be
\begin{equation}
E_j=-2\sum_{\alpha=1}^M
      [\varphi_4(z_j-t_\alpha)+\varphi_1(z_j-t_\alpha)]
\end{equation}
with the constraints $f_\alpha=0,~\alpha=1,2,\cdots,M$, where
\begin{eqnarray}
f_\alpha=-4\sum_{\beta=1,\ne\alpha}^M
       [\varphi_4(t_\beta-t_\alpha)+\varphi_1(t_\beta-t_\alpha)]
         -2\sum_{j=1}^N
       [\varphi_4(t_\alpha-z_j)+\varphi_1(t_\alpha-z_j)].
\end{eqnarray}
Here 
$\varphi_4(z)\equiv\Theta'(z)/\Theta(z)$ and
$\varphi_1(z)\equiv H'(z)/H(z)$.  $\Theta(z)$ and $H(z)$ are the Jacobi
theta-functions and they satisfy the relations
$$\sn(z)=H(z)/(\sqrt{k}\Theta(z)), \quad
\sn^2(z)+\cn^2(z)=1,\quad k^2\sn^2(z)+\dn^2(z)=1.
$$
We can write down the off-shell Bethe ansatz equations 
\begin{equation}
H_j\phi=E_j\phi+\sum_{\alpha=1}^Mxf_\alpha
     \frac{1}{\Theta(t_\alpha-z_j)H(t_\alpha-z_j)}N(t)\phi_\alpha,
\end{equation}
where $$
x=\frac{2g'(0)g(\frac{s+t}{2}+t_\alpha-z_j-\frac{1}{2})}
       {g(t_\alpha-z_j)g(\frac{s+t}{2}-\frac{1}{2})}$$
with $s,t$ being parameters and $g(u)=H(u)\Theta(u)$,
$$N(t)=c(s,t)\cdot\left(\begin{array}{cc}
  \Theta(t)H(t) & -H(t)H(t)\\
  \Theta(t)\Theta(t)& -\Theta(t)H(t)\end{array}\right).
$$
The Bethe states $\phi$ and $\phi_\alpha$ are given by
\begin{eqnarray}
\phi&=&\prod_{\alpha=1}^M\sum_{j=1}^N
       \frac{1}{\Theta(t_\alpha-z_j)H(t_\alpha-z_j)}M(t,t_\alpha,z_j)|0>,
\nonumber\\
\phi_\alpha&=&\prod_{\beta\ne\alpha}^M\sum_{j=1}^N
       \frac{1}{\Theta(t_\beta-z_j)H(t_\beta-z_j)}M(t,t_\beta,z_j)|0>,
\end{eqnarray}
where 
$$M(t,t_\beta,z_j)=\rho(s,t,t_\alpha,z_j)\cdot\left(\begin{array}{cc}
  M_{11}&M_{12}\\M_{21}&M_{22}\end{array}\right)
$$
with
\begin{eqnarray*}
&&M_{11}=2\alpha\Theta(t_\alpha-z_j)H(t+t_\alpha-z_j)
[\Theta'(t+t_\alpha-z_j)H(t+t_\alpha-z_j)\\
&&\quad\quad\quad\mbox{}-
 \Theta(t+t_\alpha-z_j)H'(t+t_\alpha-z_j)]\\
&&\quad\quad\quad\mbox{}
+\Theta(t+t_\alpha-z_j)H(t+t_\alpha-z_j)
[\Theta'(t_\alpha-z_j)H(t_\alpha-z_j)\\
&&\quad\quad\quad\mbox{}-
 \Theta(t_\alpha-z_j)H'(t_\alpha-z_j)],\nonumber\\
&&M_{12}=\frac{H'(0)}{\Theta(0)}
[\Theta^2(t+t_\alpha-z_j)H^2(t_\alpha-z_j)
 -\Theta^2(t_\alpha-z_j)H^2(t+t_\alpha-z_j)],\\
&&M_{21}=\frac{H'(0)}{\Theta(0)}
[\Theta^2(t+t_\alpha-z_j)\Theta^2(t_\alpha-z_j)
 -H^2(t+t_\alpha-z_j)H^2(t_\alpha-z_j)],\\
&&M_{22}=2\alpha\Theta(t_\alpha-z_j)H(t+t_\alpha-z_j)
[\Theta'(t+t_\alpha-z_j)H(t+t_\alpha-z_j)\\
&&\quad\quad\quad\mbox{}-
 \Theta(t+t_\alpha-z_j)H'(t+t_\alpha-z_j)]\\
&&\quad\quad\quad\mbox{}
-\Theta(t+t_\alpha-z_j)H(t+t_\alpha-z_j)
[\Theta'(t_\alpha-z_j)H(t_\alpha-z_j) \\
&&\quad\quad\quad \mbox{}-
 \Theta(t_\alpha-z_j)H'(t_\alpha-z_j)].
\end{eqnarray*}

\vskip.1in
\noindent {\bf (ii) The boundary case.}

For the boundary Gaudin model 
(\ref{8v-Gb}) we find that the eigenvalues are given by
\begin{eqnarray}
E_j^b=2\varphi(2z_j)+a\varphi(z_j+\xi)
-\sum_{\alpha=1}^M[\varphi(z_j-t_\alpha)+\varphi(z_j+t_\alpha)
\end{eqnarray}
with the constraints $f_\alpha^b=0,~\alpha=1,2,\cdots,M$, where
\begin{eqnarray}
f^b_\alpha&=&
c[\varphi(t_\alpha-\xi)+\varphi(t_\alpha+\xi)]
+2\varphi(2t_\alpha)
\nonumber\\
&&\mbox{}
+2\sum_{\beta\ne\alpha}^M
[\varphi(t_\alpha-t_\beta)+\varphi(t_\alpha+t_\beta)]
-\sum_{k=1}^N
[\varphi(t_\alpha-z_k)+\varphi(t_\alpha+z_k)].\nonumber\\
\end{eqnarray}
where $a, c$ are some complex parameters. As previous section, the
off-shell Bethe ansatz equations can be obtained by taking
$\eta\rightarrow 0$ to the 
\begin{equation}
t(z_j)\Phi^b={\cal E}(z_j)\Phi^b+\sum_{\alpha=1}^M
             F^b_\alpha\Phi_\alpha^b.
\end{equation}
Here $t(z_j),{\cal E}(z_j)$ and $F^b_\alpha$ can be found in \cite{Fan96}, 
$\Phi^b, \Phi^b_\alpha$ are Bethe states given by
\begin{eqnarray}
\Phi^b&=&B(m^0-2|t_1)\cdots B(m^0-2M|z_M)|0>,\nonumber\\
\Phi^b_\alpha&=&B(m^0-2|z_j)B(m^0-4|t_1)\cdots
        B(m^0-2\alpha|t_{\alpha-1})\nonumber\\
&&\mbox{}\times
        B(m^0-2(\alpha-1)|t_{\alpha+1})\cdots
B(m^0-2M|t_M)|0>,
\end{eqnarray}
where $m^0$ is a parameter and $B$ is the $(1,2)$ element of the
$2\times 2$ monodromy matrix which can be found in \cite{Fan96}.

\section{Summary}

We have studied the elliptic Gaudin models and solutions of the
corresponding KZ equations. In the first part of the paper, we have
constructed the Gaudin models based on the face-type elliptic
quantum groups. Hamiltonians of the models are diagonalized by using the
algebraic Bethe ansatz. With the help of these Gaudin model
Hamiltonians, we have presented two types of elliptic KZ equations and
give their solutions in the form of integrals. Then in the second part
of the paper, we have constructed and diagonalized the boundary $XYZ$
Gaudin model.

\vskip.2in
\noindent {\large\bf Acknowledgements:}

This work has been financially supported by Australian Research Council.
One of the authors (Zhao) would like to thank Professor Wei Wang for 
encouragement and the Mathematics Department of the University of
Queensland for hospitality.

\vskip.2in
\noindent {\bf Note Added:} After we finished this paper, we became
aware that the integrable representation of
solutions of the face-type elliptic KZ equations in the {\em periodic} case
have been obtained in \cite{Fel3,Tak99,Kur01} by means of techniques 
different from the ones presented in this paper.
We would like to thank T. Takebe for kindly drawing our attention to these
references and the reference \cite{Bab99}.


\vskip.3in
\begin{thebibliography}{99}
\bibitem{Gau76} M. Gaudin, J. Phys. (Paris) {\bf 37}, 1087 (1976).
\bibitem{Sei94} N. Seiberg and E. Witten, Nucl. Phys. {\bf B426}, 19
    (1994).
\bibitem{Bra99} H. Braden, A. Marshakov, A. Mirohov and A. Morozov, 
    {\em The Ruijsenaars-Schneider model in the context of
    Seiberg-Witten theory}, e-print hep-th/9902205.
\bibitem{Skl87} E.K. Sklyanin, J. Sov. Math {\bf 47}, 2473 (1989).
\bibitem{Skl96} E. K. Sklyanin and T. Takebe, Phys. Lett.   
             {\bf A219}, 217 (1996).
\bibitem{Skl99} E.K. Sklyanin, Lett. Math. Phys. {\bf 47}, 275 (1999).
\bibitem{Jur89} B. Jurco, J. Math. Phys. {\bf 30}, 1739 (1989).
\bibitem{Hik92} K. Hikami, P.P. Kulish and M. Wadati, J. Phys. Soc.
    Japan {\bf 61}, 3071 (1992).
\bibitem{Skl88} E.K. Sklyanin, J. Phys. {\bf A21}, 2375 (1988).
\bibitem{Hik95} K. Hikami, J. Phys {\bf A28}, 4997 (1995).
\bibitem{KZ} V.G. Knizhnik and A.B. Zamolodchikov, Nucl. Phys. {\bf B247},
    83 (1984).
\bibitem{Babu94} H.M. Babujian, J. Phys. {\bf A26}, 6981 (1994).
\bibitem{Hik94}  K. Hikami, J. Phys. {\bf A27}, L541 (1994).
\bibitem{Has94} T. Hasegawa, K. Hikami and M. Wadati, J. Phys. Soc. Japan,
    {\bf 63}, 2895 (1994).
\bibitem{Fei94} B. Feigin, E. Frenkel and  N. Reshetikhin,
    Commum. Math. Phys. {\bf 166}, 27 (1994).
\bibitem{fe1} G. Felder and A. Varchenko, Commun. Math. Phys.
               {\bf 181}, 741 (1996).
\bibitem{Fan97} H. Fan, B. Y. Hou and K. J. Shi,  Nucl. Phys.
    {\bf B496}, 551 (1997).
\bibitem{fe2} G. Felder and A. Varchenko, Nucl. Phys. {\bf B480},  485 (1996).
\bibitem{Hou00} B.Y. Hou, K.J. Shi, R.H. Yue and S.Y. Zhao, Northwest
    University preprint, 2000.
\bibitem{Bax} R.J. Baxter, Exactly solved Models in Statistical Mechanics 
    (Academic Press, Londom, 1992).
\bibitem{Bab99} H. Babujian, A. Lima-Santos and R.H. Poghossian,
    Interna. J. Mod. Phys. {\bf A14}, 615 (1999).
\bibitem{inami} B.Y. Hou and R.H. Yue, Phys. Lett. {\bf A183}, 169
    (1993); T. Inami and H. Konno, J. Phys. {\bf A27}, L913 (1994).
\bibitem{Fan96} H. Fan, B.Y. Hou, K.J. Shi and Z.X. Yang, Nucl.Phys. 
    {\bf B478},  723 (1996).
\bibitem{Fel3} G. Felder and A. Varchenko, Internat. Math. Res. Notices,
    {\bf 5}, 221 (1995).
\bibitem{Tak99} T. Takebe, Communi. Math. Phys. {\bf 204}, 587 (1999).
\bibitem{Kur01} G. Kuroki and T. Takebe, J. Phys. {\bf A34}, 2403
    (2001).
\end{thebibliography}
\end{document}